# One-dimensional nature of the magnetic fluctuations in YBa$_2$Cu$_3$O$_{6.6}$

H. A. Mook\*, Pengcheng Dai\*, F. Doğan† & R. D. Hunt\*

\* *Oak Ridge National Laboratory, Oak Ridge, Tennessee 37831, USA*
† *Department of Materials Sciences and Engineering, University of Washington, Seattle, Washington 98195, USA*

**There is increasing evidence that inhomogeneous distributions of charge and spin—so-called 'striped phases'—play an important role in determining the properties of the high-temperature superconductors. For example, recent neutron-scattering measurements on the YBa$_2$Cu$_3$O$_{7-x}$ family of materials show both spin and charge fluctuations that are consistent with the striped-phase picture. But the fluctuations associated with a striped phase are expected to be one-dimensional, whereas the magnetic fluctuations observed to date appear to display two-dimensional symmetry. We show here that this apparent two-dimensionality results from measurements on twinned crystals, and that similar measurements on substantially detwinned crystals of YBa$_2$Cu$_3$O$_{6.6}$ reveal the one-dimensional character of the magnetic fluctuations, thus greatly strengthening the striped-phase interpretation. Moreover, our results also suggest that superconductivity originates in charge stripes that extend along the** b **crystal axis, where the superfluid density is found to be substantially larger than for the** a **direction.**

Striped phases are inhomogeneous distributions of charge and spin that have been suggested to account for many of the unusual properties of the high-$T_c$ copper oxide superconductors[1-7] ($T_c$ is the superconducting transition temperature). In the simplest picture, the charge and spin can be thought of as being confined to separate linear regions in the crystal and thus resembling stripes. We expect static striped phases for the YBa$_2$Cu$_3$O$_{7-x}$ materials not to coexist with superconductivity, except perhaps for materials with low oxygen contents and low values of $T_c$. However, neutron-scattering measurements have shown results for fluctuations of both spin[8,9] and charge[10] that support the existence of a dynamic striped phase in YBa$_2$Cu$_3$O$_{7-x}$ materials that have high $T_c$ values. Nevertheless, a difficulty for the dynamic striped-phase picture is that, for both YBa$_2$Cu$_3$O$_{7-x}$ (refs 8, 9) and La$_{2-x}$Sr$_x$CuO$_4$ (refs 11, 12) copper oxide superconductors, the magnetic fluctuations stemming from the spins have displayed a four-fold pattern at incommensurate points around the magnetic (1/2, 1/2) reciprocal lattice position, as shown in Fig. 1A. The interpretation of these measurements has been unclear, with both striped phases[1-7] and nested Fermi surfaces[13-15] being possible explanations. Because the striped phase is expected to propagate along a single direction, only a single set of satellites around the antiferromagnetic position should be observed. Tranquada[7,16] has suggested that the four-fold symmetry might arise from stripes that alternate in direction as the planes are stacked along the **c** axis.

The problem in interpreting the four-fold pattern stems from the fact that the crystals used in the neutron experiments are twinned, so that along a given **a** or **b** direction, domains with lattice spacing $a$ or $b$ exist in equal proportion. This makes it impossible to distinguish whether satellites originate from the **a**\* or **b**\* direction in the reciprocal lattice. What is needed to resolve the problem of interpreting the magnetic measurements is either an untwinned crystal, or a crystal with a substantial difference in the domain population along the different directions in the **a**–**b** plane.

## Sample preparation

We have attempted to detwin YBa$_2$Cu$_3$O$_{7-x}$ crystals by applying a high pressure to one of the **a**(**b**) directions of a crystal during oxygenation, so that as the crystal undergoes the tetragonal to orthorhombic transition, the squeezed direction prefers to be the lower-lattice-constant **a** direction. Crystals are commonly detwinned in this manner, but it is challenging to accomplish this with a crystal large enough to observe the magnetic incommensu-rate scattering with neutrons. Typically, either the crystal breaks or only a small difference in domain population is obtained. The 4.39-gram crystal used in the neutron measurements was clamped vertically, with the **c** axis perpendicular to the load direction, between two aluminium oxide ceramic rods that were attached to the water-cooled plates of a hydraulic press. A thin layer of refractory wool was placed between the sample and rods to reduce any possible non-uniform stress distribution under applied pressure, and to enhance oxygen diffusion along the clamped surfaces. Oxygenation of the sample took place in a tube furnace at 400 °C while a constant pressure of 13.9 MPa was applied for 120 hours in a flowing-oxygen gas environment. The sample was then reduced to an oxygen content of about 6.6, in accordance with the techniques developed during an extensive experimental study[17] on the nonstoichiometry in YBa$_2$Cu$_3$O$_{7-x}$ (an oxygen content of 6.6 indicates that $x$ in this formula is 0.4). An equation was developed in this study relating the oxygen content and $T_c$ to the partial oxygen pressure and temperature. The detwinning process resulted in a sample in which the domain population has roughly a 2:1 ratio.

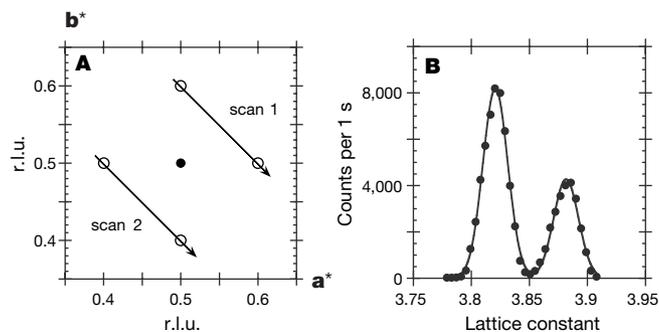

**Figure 1** Diagram showing the scans used in the experiment, and a plot of the twin domain population. **A**, Open circles show the positions in reciprocal lattice units (r.l.u.) of the magnetic incommensurate satellites that surround the antiferromagnetic position shown by the filled circle. The arrows show the two scans used in the experiment. The centre of scan 1 is at (0.55, 0.55, 2) in r.l.u., while the centre of scan 2 is at (0.45, 0.45, 2) r.l.u. **B**, Scan for the **a**\* direction obtained from the (2, 0, 0) reflection given in units of the lattice constant for the partially detwinned sample. The lattice constants are those expected for an oxygen concentration near 6.6. The peak heights give the domain population of the partially detwinned sample.

The Bragg scattering from one of the **a**(**b**) directions from the crystal is shown in Fig. 1B. The pattern was taken with the HB-3 spectrometer at the High-Flux Isotope Reactor (HFIR) in Oak Ridge, Tennessee, USA, using 14.7-meV neutrons obtained from the (002) reflection of a Be monochromator. The angular collimation used was 10 minutes before and after the sample. The peak corresponding to lattice constant $a$ is considerably bigger than that for the lattice constant $b$, so we denote this face as the **a** face of the crystal. Least-squares fitting a gaussian distribution to the scattering indicates a domain ratio ($a/b$) of $1.92 \pm 0.05$. Measurements on the **b** face confirmed this result with an identical domain ratio for ($b/a$) within the error of the measurement.

### The pattern of the magnetic fluctuations

We then modified the spectrometer configuration so that we could observe the incommensurate magnetic fluctuations for both the **a*** and **b*** directions in the reciprocal lattice of the crystal. Pyrolytic graphite monochromator and analyser crystals were installed on the HB-3 triple-axis spectrometer, and the spectrometer was set for an energy transfer of 24 meV using a final energy of 30.5 meV. Collimations were 48-40-60-120 minutes from in front of the monochromator until after the analyser. Measurements were first made on the 25.6-gram twinned crystal used to originally determine the location of the incommensurate scattering peaks[9]. The crystal was oriented so that the scans through the satellites could be made as shown in Fig. 1A while keeping **c*** set at 2 r.l.u. (reciprocal lattice units) so as to maximize the bilayer structure factor[8,9]. These scans have the advantage that the neutron detector remains at a nearly constant angle, thus avoiding background contamination from the direct neutron beam of the triple-axis spectrometer. This permits a measurement at the small total momentum transfer used, which maximizes the magnetic form factor while minimizing the effects of phonon scattering. For $YBa_2Cu_3O_{6.6}$ the satellites are found in the twinned crystal at approximately 0.1 r.l.u. along **a*** and **b*** from the magnetic superlattice peak. The scan shown in Fig. 2A is denoted scan 1 in Fig. 1A. The centre of this scan is at 0.55 r.l.u. while the peaks are expected at 0.5 and 0.6 r.l.u. along **a***. The scan in Fig. 2B is denoted scan 2 in Fig. 1A. The scan centre is at 0.45 r.l.u. in this case, with the peaks expected at 0.4 and 0.5 r.l.u. along **a***. Least-squares fitting of gaussian distributions yields a ratio of the intensities of the satellites along **a*** to those along **b*** of $1.05 \pm 0.05$, which is consistent with the equal intensities expected for the two directions expected for the completely twinned crystal.

Measurements for the partially detwinned sample are shown in Fig. 2C, D. Least-squares fitting in the same manner for these data results in a ratio of the satellite intensities along **a*** to those along **b*** of $1.95 \pm 0.21$, which is the same within the error bars as the population of the domains for the two directions. Thus, it follows within an error of 11% that 100% of the satellites stem only from the **a*** domain, showing that the spin fluctuations have a one-dimensional (1D) modulation along **a**. A 1D static magnetic modulation along **b** has recently been found by Wakimoto *et al.*[18] for the insulating material $La_{1.95}Sr_{0.05}CuO_4$. This suggests the possibility of a static striped phase propagating along **b**, although in this case no evidence of charge scattering was identified.

### Phonon measurements

Now that we have discovered that the incommensurate magnetic scattering has a 1D modulation vector for $YBa_2Cu_3O_{6.6}$, it is worthwhile revisiting the measurements of the charge scattering as imaged by the phonons[10]. Because the striped-phase modulation vector occurs along **a**, the phonons from the **b** direction should display normal behaviour. In a twinned crystal, the charge scattering measured in the **a***(**b***) direction should have one component that has normal behaviour and one component that is strongly affected by the striped phase. In fact, the measured line shapes shown in figure 2 of ref. 10 give an indication that the wide peak where the charge anomaly is found may consist of two parts. One of the parts would be expected to be sharp and occur at the unperturbed phonon position, and the other would be strongly damped by the charge fluctuations. Calculations of the phonon line shapes in striped-phase materials show a rather complicated line shape resulting from the interaction of the charge fluctuations with the phonons (N. Wakabayashi *et al.*, manuscript in preparation). Nevertheless, the knowledge that the striped-phase modulation vector occurs mostly along **a** justifies a higher-resolution study of the phonon anomaly.

Figure 3 shows a high-resolution measurement at the 4.25 r.l.u. position for the optical phonon studied in ref. 10 for $YBa_2Cu_3O_{6.6}$

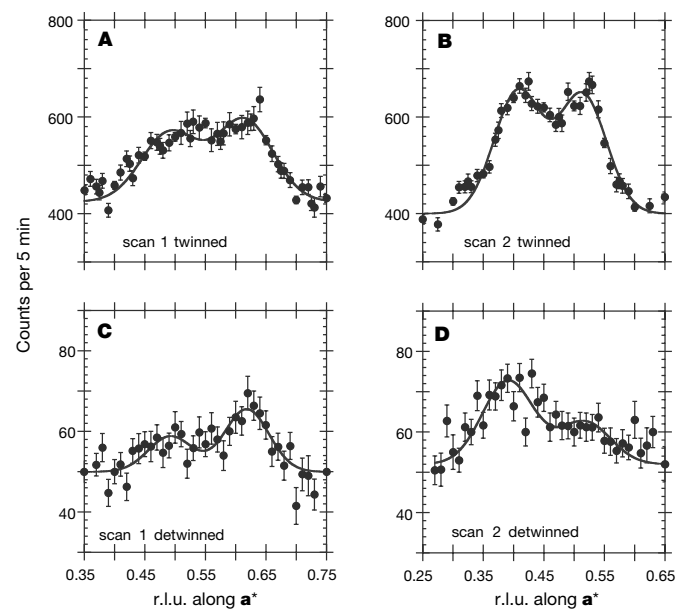

**Figure 2** Scans of the magnetic incommensurate scattering for the twinned and partially detwinned sample taken at 10 K. The scan directions are shown in Fig. 1A. The counting errors shown were obtained by averaging a number of runs together. The larger magnetic scattering for scan 2 is accounted for by the Cu magnetic form factor. The scan is made in the diagonal direction shown, with the scale on the graphs giving the component along **a***.

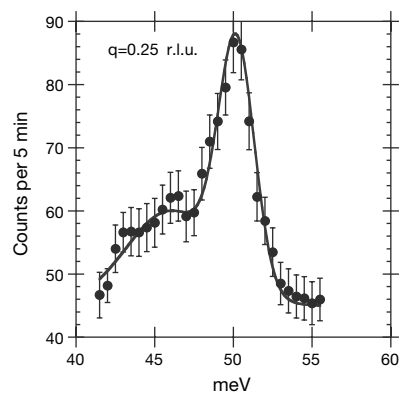

**Figure 3** Optical phonon measurements on $YBa_2Cu_3O_{6.6}$. The figure shows a high-resolution measurement at 10 K of the optical phonon discussed in ref. 10. The measurement was made to image the charge fluctuations in $YBa_2Cu_3O_{6.6}$. The higher resolution makes it clearer that the broad line shape resulting from the charge fluctuations can be interpreted to consist of the superposition of two lines; a normal narrow line from the **b** twin domain, and a broad line from the **a** twin that contains the striped-phase 1D modulation vector.

using the twinned crystal. This is in the region of momentum space where we expect the phonons to be strongly affected by the striped phase. The fit shown is for two gaussian distributions. The position of the narrow peak occurs at 50.1 meV, consistent with the value found in lightly doped materials where any striped phase would have a much smaller wavevector, while the width of 2.9 meV is the same as the experimental resolution. The other distribution is found to have a wide width of 5.7 meV, as might be expected from the interaction with a stripe-phase dynamic charge density wave. The areas under the gaussian distributions are equal within experimental errors. This is consistent with our result for the magnetic scattering, and indicates a method of determining whether the charge fluctuations are 1D or 2D.

### Stripes versus Fermi liquid

The Fermi liquid concept, with its concomitant quasiparticles and Fermi surface, has been successful in solving many of the important problems in condensed-matter physics, including conventional superconductivity. This approach for the layered copper oxide superconductors results in a 2D Fermi surface. We are unaware of any Fermi surface calculation that can account for our measured results unless 1D effects are somehow incorporated into the calculation. The stripe model takes a completely different approach, wherein the electronic matter self-organizes into a topologically 1D-quantum texture. Such a concept naturally incorporates the 1D modulation observed in the present experiment. This provides compelling evidence that the observed spin and charge fluctuations stem from a dynamic striped phase. We do not have sufficient information to understand in detail the characteristics of such a phase. Nevertheless, such a phase is a novel state of matter decidedly different from those stemming from the conventional Fermi-liquid ideas.

However, the striped-phase model is not able to completely characterize the observed magnetic fluctuations. Much of the weight of the total magnetic scattering is found in the commensurate resonance excitation[19–21]. So far, it is unclear how the striped-phase concept can account for the resonance. It is also uncertain if the striped phase has any direct connection to the superconductivity. It may be, though, that the 1D nature of the striped phase simplifies the difficulties encountered in understanding the nature of the superconducting pairing mechanism[22]. The charge stripes extend along **b**, which is in the same direction as the Cu–O chains found in the $YBa_2Cu_3O_{7-x}$ structure. If superconductivity occurs in the charge stripes, anisotropic superconducting behaviour should be found in the **a–b** plane. Far-infrared spectroscopy measurements by Basov et al.[23,24] show that the London penetration depth is considerably smaller, or the superfluid density larger, along the **b** direction for the fully doped material. This was attributed to a substantial portion of the superconducting condensate residing on the chains. However, the superfluid density was also found to be about 1.7 times larger for the underdoped material $YBa_2Cu_3O_{6.6}$, where the missing atoms in the chains would be expected to inhibit the chain contribution to the superconductivity. The present measurements show that the stripes extend in the correct direction to account for the observed anisotropy in the superfluid density. □

**Acknowledgements**

The work at Oak Ridge National Laboratory was supported by the US Department of Energy.

Correspondence and requests for materials should be addressed to H.A.M. (e-mail: ham@ornl.gov).